\documentclass[twocolumn,amsfonts,showpacs,superscriptaddress,nofootinbib]{revtex4-1}
\usepackage{amsfonts,amssymb,amsmath}
\usepackage{lipsum}
\usepackage{mathtools}
\usepackage[utf8]{inputenc}
\usepackage[pdftex,breaklinks=true,colorlinks=true]{hyperref}
\usepackage[pdftex]{graphicx}
\usepackage{epstopdf}
\usepackage{leftidx}
\graphicspath{{Pictures/}}

\newcommand{\bc}{\begin{center}}
\newcommand{\ec}{\end{center}}
\def\ba#1{\begin{array}{#1}\displaystyle}
\newcommand{\ea}{\end{array}}

\newcommand{\beq}{\begin{equation}}
\newcommand{\eeq}{\end{equation}}
\newcommand{\beqa}{\begin{eqnarray}}
\newcommand{\eeqa}{\end{eqnarray}}

\newcommand{\n}{\nonumber\\}
\newcommand{\bi}{\begin{itemize}}
\newcommand{\ei}{\end{itemize}}

\newcommand{\p}{\partial}

\newcommand{\dd}{{\rm d}}

\newcommand{\dr}{\mathrm{dr}}

\newcommand{\ttbar}{$T\bar{T}$}

\newcommand{\eff}{{\rm eff}}

\def\eqref#1{(\ref{#1})}

\newcommand{\cL}{{\cal L}}
\newcommand{\bcL}{\bar{\cal L}}

\usepackage{amsmath}	
\begin{document}


\title{$T\bar{T}$-deformed conformal field theories out of equilibrium}

\author{Marko Medenjak}
\affiliation
{Institut de Physique Th\'eorique Philippe Meyer, \'Ecole Normale Sup\'erieure, \\ PSL University, Sorbonne Universit\'es, CNRS, 75005 Paris, France}

\author{Giuseppe Policastro}
\affiliation{Laboratoire de Physique de l’Ecole Normale Sup\'erieure, CNRS, Universit\'e PSL, Sorbonne 
Universit\'es, Universit\'e Pierre et Marie Curie, 24 rue Lhomond, 75005 Paris, France}

\author{Takato Yoshimura}
\affiliation{Department of Physics, Tokyo Institute of Technology, Ookayama 2-12-1, Tokyo 152-8551,
Japan}
\affiliation
{Institut de Physique Th\'eorique Philippe Meyer, \'Ecole Normale Sup\'erieure, \\ PSL University, Sorbonne Universit\'es, CNRS, 75005 Paris, France}

\begin{abstract}
We consider the out-of-equilibrium transport in \ttbar-deformed (1+1)-dimension conformal field theories (CFTs). The theories admit two disparate approaches, integrability and holography, which we make full use of in order to compute the transport quantities, such as the the exact non-equilibrium steady state currents. We find perfect agreements between the results obtained from these two methods, which serve as nontrivial checks of the $T\bar{T}$-deformed holographic correspondence from the dynamical standpoint. It turns out that integrability also allows us to compute the momentum diffusion, which is given by a universal formula. We also remark on an intriguing connection between the $T\bar{T}$-deformed CFTs and reversible cellular automata.
\end{abstract}

\maketitle
\noindent {\bf\em Introduction.}
Understanding the out-of-equilibrium phenomena of many-body quantum systems represents one of the great challenges of modern physics. Solving a full dynamics is typically unfeasible due to the many-body nature of the problem, however, for certain systems we can access the hydrodynamic regimes analytically. Such systems can therefore provide precious insights about the effects of microscopic details on the large scale dynamics.

Recently, we have witnessed an evolution of two particularly successful theoretical approaches in the study of out-of-equilibrium phenomena; holography and generalized hydrodynamics (GHD). Holography connects strongly-coupled conformal field theories and weakly coupled gravitational theories on the  anti-de Sitter space (AdS) background \cite{Maldacena:1997re,Witten:1998qj,Gubser:1998bc}, and GHD provides the hydrodynamic description of integrable models \cite{PhysRevX.6.041065,PhysRevLett.117.207201}. Both of the theories led to a myriad of new results on transport and hydrodynamics \cite{Policastro:2001yc,Kovtun:2004de,Son:2009tf, CasalderreySolana:2011us, SciPostPhys.3.6.039,PhysRevLett.121.160603,De_Nardis_2019}, however, no connection between the two approaches was established to date. In this letter we will make such a connection, by studying the out-of-equilibrium properties of $T\bar{T}$-deformed CFTs \cite{Zamolodchikov:2004ce,Dubovsky2012,Caselle2013,Cavagli2016,SMIRNOV2017363,Kraus2018,Cardy2018,Conti2019,jiang2019lectures,cardy2020toverline}.

\begin{figure}[h!]
\centering
\includegraphics[width=8.5cm]{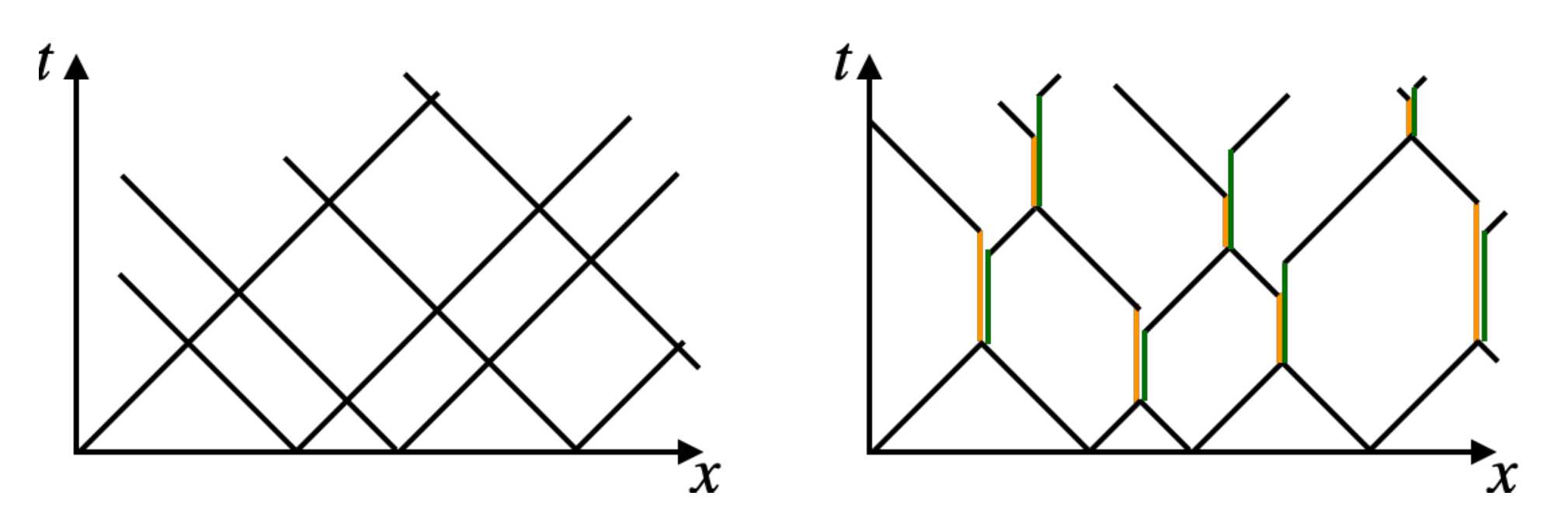}
\caption{Trajectories of right/left movers in a pure CFT and a \ttbar-deformed CFT on a hydrodynamic scale. The trajectories of incoming particles are the same, but $R-L$ scatterings induced by the \ttbar-deformation amount to time delays. Note that, during each scattering, the time delays are generically different for right and left movers.}
\label{ttbar_scattering}
\end{figure}

$T\bar{T}$-deformation is an irrelevant deformation, which makes it \emph{especially relevant} when considering out-of-equilibrium properties of gapless systems at low temperatures. Namely, the departure from the low energy spectrum described by CFTs, is governed by irrelevant perturbations, leading to the non-linear Luttinger liquid theory \cite{Imambekov228,1367-2630-17-10-103003}. $T\bar{T}$-deformation represents a perfect testbed for studying the effects of such perturbations (see Fig.\,\ref{ttbar_scattering}). Firstly, the theory is exactly solvable, and secondly, despite of being away from criticality, a dual theory of gravity has been conjectured. Note also that the composite operator $T\bar{T}$ is in general the leading correction around the IR fixed point in cases where the perturbation does not break translational invariance (i.e. no Umklapp scattering).
In this letter we will consider two out-of-equilibrium aspects of $T\bar{T}$-deformed CFTs. First, we will look at the partitioning protocol, where the two sides of the system are prepared in a thermal state at different temperatures and provide a closed-form expression for the nonequilibrium steady state (NESS), which emerges after the equilibrium is established locally. Secondly, we will analytically calculate the energy and momentum Drude weights. Surprisingly, we find that both NESS and Drude weights are universal, in the sense that they depend solely on the central charge $c$ and on the deformation strength, and not on other details of the underlying CFT. Importantly, we show that the results obtained through holography perfectly match with the GHD prediction. This confirms the universality of the results, showing that they apply beyond the realm of conventional integrable models.
Going beyond the Euler scale physics, we also show that the momentum diffusion, which is absent in a pure CFT, is given by a universal formula. 
Finally, we notice a curious relation between the $T\bar{T}$-deformed CFTs and the Rule 54 cellular automaton (RCA 54) \cite{Bobenko1993}.

\vspace{0.2cm}

\vspace{0.2cm}

\noindent {\bf\em Hydrodynamics of \ttbar-deformed CFTs.}\quad In (1+1)-dimensional quantum field theory, $T\bar{T}$-deformation arises as an infinitesimal deformation to the Lagrangian $\mathcal{L}^{(\sigma)}$ by the composite operator $\mathtt{T}\bar{\mathtt{T}}=T^{0\mu}T^{1\nu}\epsilon_{\mu\nu}$: $\mathcal{L}^{(\sigma+\delta\sigma)}=\mathcal{L}^{(\sigma)}-\frac{\delta\sigma}{2\pi^2}\mathtt{T}\bar{\mathtt{T}}$.
Note that $T_{\mu\nu}$ depends on the value of $\sigma$. If we can obtain a solution of the undeformed theory, we can also solve its deformation, since it satisfy the Burgers equation  \cite{Zamolodchikov:2004ce}
\begin{equation}\label{burgers}
    \p_\sigma E_n(R,\sigma)=E_n(R,\sigma)\p_RE_n(R,\sigma)+\frac{1}{R}P^2_n(R),
\end{equation}
where $R$ is the volume of the (compactified) system and $E_n(R,\sigma=0)$ the energies of an undeformed theory. Notice that the momenta $P_n(R)=2\pi p_n/R,\,p_n\in\mathbb{Z}$ remain undeformed. 

Here, we will focus on \ttbar-deformed  CFTs which pertain a thermodynamic description in terms of two non linear integral equations (NLIE) only, one for the left movers and one for the right movers. The description of CFTs by NLIE has been developed for many models, including the minimal models \cite{Bazhanov1996} and Liouville CFTs \cite{zamolodchikov2012quantum}, and turns out to be particularly useful for our purpose. 
These NLIEs provide pseudo energies of the quasiparticles comprising the state \cite{comment}. The NLIEs governing the pseudoenergies $\varepsilon_\pm(\theta)$ of CFTs in a boosted state
$\rho\sim e^{-\beta H+\nu P}$, where $H$ and $P$ are the hamiltonian and momentum operator, read
\begin{equation}\label{nlieGGE1}
    \varepsilon_\pm(\theta)=\beta_\pm E_\pm(\theta)-T\star L_\pm(\theta),
\end{equation}
where $\beta_\pm=\beta\mp\nu$, $E_\pm(\theta)=Me^{\pm\theta}/2$, and $L_\pm(\theta)=\log(1+e^{-\varepsilon_\pm(\theta)})$. Here $\star$ stands for the convolution: $T\star L(\theta)=\int\dd\theta'T(\theta,\theta')L(\theta')$, and the phase shift $T(\theta)$ depends on the model parameter.
Note that the two equations \eqref{nlieGGE1} are decoupled if \ttbar-deformation is not present.
In thermal equilibrium (i.e. $\nu=0$), the central charge of the system can be read off from the free energy $\mathtt{f}=\mathtt{f}_++\mathtt{f}_-$ as $c=-6\beta \mathtt{f}/\pi$, where
\begin{equation}\label{centralcharge}
     \mathtt{f}_\pm=-\int_\mathbb{R}\frac{\dd\theta}{2\pi}p'_\pm(\theta)L_\pm(\theta).
\end{equation}
where $p_\pm(\theta)=\pm Me^{\pm\theta}/2$ is the bare momentum. Notice that in the presence of higher charges the free energy is no longer related to $c$.

Having the NLIEs for CFTs at our disposal, we are in a position to turn on the \ttbar-deformation, which induces the $R-L$ scatterings, resulting in the phase shift $\tilde{T}_{\pm\mp}(\theta,\theta')=-\sigma p_+(\theta)p_-(\theta')/(2\pi)$. 
The TBA equations for the \ttbar-deformed CFT then reads
\begin{equation}
    \varepsilon_\pm(\theta)=\beta_\pm E_\pm(\theta)-T\star L_\pm(\theta)-\tilde{T}_{\pm\mp}\star L_\mp(\theta),
\end{equation}
which can also be written, by noting \eqref{centralcharge}, as
\begin{equation}\label{nlieGGE2}
    \varepsilon_\pm(\theta)=\left(\beta_\pm+\sigma\mathtt{f}_\mp\right)E_\pm(\theta)-T\star L_\pm(\theta).
\end{equation}
Observing that the NLIEs are exactly the same as \eqref{nlieGGE1} with the replacement $\beta_\pm\mapsto\tilde{\beta}_\pm=\beta_\pm+\sigma\mathtt{f}_\mp$, we can recast the NLIEs into a more transparent way \cite{Cavagli2016} $ \varepsilon_\pm(\theta)=\tilde{\beta}_\pm(\beta,\nu) E_\pm(\theta)-T\star L_\pm(\theta),$
where
\begin{equation}
    \tilde{\beta}_\pm(\beta,\nu)=\frac{\beta_\pm}{2}\left(1+\sqrt{1-\frac{\pi\sigma c}{3\beta_+\beta_-}}\right).
\end{equation}
In integrable systems, the bare energies and momentum are renormalized due to the interaction. The effect of renormalization is conveniently encoded through the ``dressing''. In general, when the phase shift $T_{ab}(\theta,\theta')$ is symmetric (with respect to $(a,\theta)\in\{\pm\}\times\mathbb{R}$) the dressing operation to a function $f_\pm(\theta)$ is defined by $f^\dr_\pm(\theta)=f_\pm(\theta)+[T\star n_\pm f_\pm^\dr](\theta)+[\tilde{T}\star n_\mp f_\mp^\dr](\theta)$. In GHD, we in particular need $E_\pm^\dr,p^\dr_\pm$ and $(E'_\pm)^\dr,(p'_\pm)^\dr$. Note that $E_\pm^\dr=(p'_\pm)^\dr$ and $p^\dr_\pm=(E'_\pm)^\dr$, since $E_\pm=p'_\pm$ and $p_\pm=E'_\pm$, which is a unique property in massless systems. Using $E^\dr_\pm(\theta)=\p \varepsilon_\pm(\theta)/\p\beta$ and  $p^\dr_\pm=-\p \varepsilon_\pm(\theta)/\p\nu$, we can imediately obtain 
\begin{align}
    (p'_\pm)^\dr(\theta)&=\p_\beta\tilde{\beta}_\pm E_\pm(\theta)+[T\star n(p'_\pm)^\dr](\theta)\n
   (E'_\pm)^\dr(\theta)&=-\p_\nu\tilde{\beta}_\pm E_\pm(\theta)+[T\star n(E'_\pm)^\dr](\theta),
\end{align}
 from \eqref{nlieGGE2}.
Crucially, this implies that the group velocity of excitations in our theory does not depend on the rapidity variable: $ v^\eff_\pm(\theta)=(E'_\pm)^\dr(\theta)/ (p'_\pm)^\dr(\theta)=-\p_\nu\tilde{\beta}_\pm/\p_\beta\tilde{\beta}_\pm$.
 In particular in a thermal state it takes a rather simple form
 \begin{equation}\label{effective-v}
     v^\eff_\pm(\theta)=\pm\sqrt{1-\frac{\pi\sigma c}{3\beta^2}}=:\pm v^\eff(\beta).
 \end{equation}
The independence of the effective velocities on $\theta$ is one of the  main features of the \ttbar-deformed CFTs, which has important consequences on physical properties and also allows us to obtain the closed form analytical results. 
For this to be true, it is crucial that the phase shift $\tilde{T}$ is proportional to $p_+(\theta)p_-(\theta')$, and the  masslessness of excitations of the system itself is not sufficient \cite{Horvth2019}. The  expression for effective velocities can be recast into a more suggestive form. Recall that the energy densities and the energy currents can be obtained from the free energy \eqref{centralcharge}: $\rho_\pm=\p \mathtt{f}_\pm/\p\beta$ and $\rho_\pm v^\eff_\pm=-\p \mathtt{f}_\pm/\p\nu$ respectively, where $\rho_\pm$ are chiral energy densities $\rho_\pm=\int\frac{\dd\theta}{2\pi}p'_\pm(\theta)n_\pm(\theta)E^\dr_\pm(\theta)$, we find
 \begin{equation}\label{eff-velocity}
     v^\eff_\pm=\frac{\pm1+\sigma(\rho_+-\rho_-)}{1+\sigma(\rho_++\rho_-)}.
 \end{equation}
Importantly, using the GHD equation $\p_t(p'_\pm)^\dr(\theta)+\p_x(v^\eff_\pm(p'_\pm)^\dr(\theta))=0$, this implies that the hydrodynamic equation for the energy densities turns out to be a closed equation
\begin{equation}\label{RCA54}
\p_t\rho_\pm+\p_x(v^\eff_\pm\rho_\pm)=0.
\end{equation}
This is in stark contrast with the hydrodynamics of typical integrable systems: hydrodynamic equations for different charges are typically highly intertwined, and in order to predict their dynamics, we have to solve all of the equations self-consistently.

Surprisingly, if we define new chiral densities $\rho_{\mathrm{sol},\pm}:=\sigma\rho_\pm$, the equations become the GHD equations for solition densities $\rho_{\mathrm{sol},\pm}$ in an integrable cellular automaton model called Rule 54 chain (RCA 54) \cite{Friedman2019}.
RCA 54 is one of the simplest interacting systems \cite{Bobenko1993}, comprising the left and the right moving solitons propagating at a constant velocity $\pm 1$, which incur a phase shift by a unit lattice spacing upon scattering. According to the above observation, we have a dictionary which connects the quantities in the \ttbar-deformed CFTs and in the RCA 54: the energy densities in \ttbar-deformed CFTs, $\rho_\pm$, can be interpreted as the densities of right/left moving solitons in the RCA 54. The energy current conservation $\rho_+-\rho_-$ in the \ttbar-deformed CFTs is reflected by the conservation of the particle current is in the RCA 54. Note however that thermodynamics of the two models differ due to the hard-core repulsion between the same species in the RCA 54, which is absent in \ttbar-deformed CFTs. 
\vspace{0.2cm}

\noindent {\bf\em Exact non-equilibrium steady states (NESSs) in $T\bar{T}$-deformed CFTs.}\quad 
To decipher the structure of NESSs in $T\bar{T}$-deformed CFTs, we study the partitioning protocol starting from two thermal baths with temperatures $T_R$ and $T_L$. To be more precise, we choose the initial condition $\rho_\pm(x,0)=\rho_{\pm,L}\vartheta(-x)+\rho_{\pm,R}\vartheta(x)$, where $\rho_{\pm,R/L}$ are energy densities evaluated with respect to the temperatures $T_{R/L}$.  A Riemann problem of \eqref{RCA54} was in fact solved before in the scope of the hydrodynamics for two beams of KdV solitons \cite{PhysRevLett.95.204101}. The solution is given by two contact discontinuities (i.e. shocks that do not accompany entropy production) along $\xi=-v^\eff_L$ and $\xi=v^\eff_R$, where $\xi=x/t$ and $v^\eff_{R,L}=v^\eff(\beta_{R,L})$ \cite{medenjak2020thermal}. This implies that the fluid profile in a \ttbar-deformed CFT is quite similar to that of a pure CFT; a current-carrying steady state emerges inside the light cone $-v^\eff_R<\xi<v^\eff_L$, while no current is present outside. The energy and momentum NESS currents $\langle j_E\rangle=\rho_+-\rho_-$ and $\langle j_P\rangle=\rho_+v^\eff_+-\rho_-v^\eff_-$ inside of the light cone take a simple form
\begin{align} \label{cicurrent2}
    \langle j_E\rangle_\mathrm{NESS}&=\frac{\pi c }{12}e_{RL}\left(\tilde{T}^2_L-\tilde{T}^2_R\right),\n
    \langle j_P\rangle_\mathrm{NESS}&=\frac{\pi c }{12}e_{RL}\left(\tilde{T}^2_L+\tilde{T}^2_R-\frac{\pi c\sigma}{6}\tilde{T}^2_L\tilde{T}^2_R\right),
\end{align}
where $ e_{RL}=1/(1-\left(\frac{\pi\sigma c}{12}\right)^2\tilde{T}^2_L\tilde{T}^2_R)$
and $\tilde{T}_{R,L}=1/\tilde{\beta}_\pm(1/T_{R,L},0)$. In the limit $\sigma\to0$ \eqref{cicurrent2} reduces to a pure Stefan-Boltzmann law, which is a consequence of the conformal invariance \cite{bernard2012energy}.  Since \ttbar-deformation induces the $R-L$ interaction,  the NESS currents can no longer be written as $f(\beta_L)- f(\beta_R)$.  Importantly, the mixing of the right and the left movers in \eqref{cicurrent2}, is absent in the first order in $\sigma$, confirming the perturbative results in \cite{Bernard_2016}. 
We can also infer  that NESS corresponds to the boosted thermal state with the effective temperature $T=\sqrt{T_RT_L}$ and the boost parameter  $\tanh\nu=(\beta_L-\beta_R)/(\beta_L+\beta_R)$, which matches precisely with the structure of the NESS in the undeformed CFTs. 
 
Having obtained exact NESS currents, we proceed to compute the energy and momentum Drude weights, which are defined by
\begin{equation}
    D_{ij}=\lim_{t\to\infty}\int_{-t}^t\frac{\dd s}{2t}\int_\mathbb{R}\dd x\langle j_i(x,s)j_j(0,0)\rangle^c,
\end{equation}
where $j_i(x,t)$ is the current associated with the conserved charge density $q_i$ via the continuity equation $\p_t q_i+\p_x j_i=0$, and ``$c$'' refers to the connected correlation function. To do so, we make use of the general formula established within the framework of GHD \cite{SciPostPhys.3.6.039},
which leads to the following neat expressions
\begin{equation} \label{Drude-weights}
    D_{EE}=\frac{\pi c}{3\beta^3}\frac{1}{v^\eff},\quad D_{PP}=\frac{\pi c}{3\beta^3}v^\eff.
\end{equation}
Remarkably, these formulae have exactly the same form as in pure CFTs, after the light-cone velocity is appropriately modified $v=1\mapsto v^\eff$ \cite{bernard2012energy}. In the next section, we will exactly reproduce all of the GHD results obtained so far, by re-formulating the problem in the language of AdS/CFT correspondence.

But before we move to holography let's discuss the corrections to the purely ballistic transport. These corrections are typically governed by diffusive broadening of  the ballistic trajectories, which is characterized by the diffusion constants $\mathfrak{D}_i^{\,\,j}$ that is obtained from the Kubo formula $\mathfrak{L}=\mathfrak{D}C$, where $\mathfrak{L}$ is the Onsager matrix
\begin{equation} \label{Onsager}
    \mathfrak{L}_{ij}=\int_\mathbb{R}\dd t\left(\int_\mathbb{R}\dd x\langle j_i(x,t)j_j(0,0)\rangle^c-D_{ij}\right)
\end{equation}
and $C_{ij}=\int_\mathbb{R}\dd x\langle q_i(x,0)q_j(0,0)\rangle^c$ the susceptibility matrix. A general expression for diffusion was derived in the scope of GHD \cite{PhysRevLett.121.160603,De_Nardis_2019}. First of all the energy diffusion $\mathfrak{L}_{EE}$ vanishes due to the Lorentz invariance. To our surprise, the momentum diffusion $\mathfrak{L}_{PP}$ admits a universal formulae as well
\begin{equation} \label{LPP}
  \mathfrak{L}_{PP}=\frac{\sigma^2}{2}v^\eff D_{EE}^2=\frac{\sigma^2}{2}\frac{D_{PP}^2}{(v^\eff)^3},
\end{equation}
 Observe that the leading term is quadratic in $\sigma$. It is astonishing that the momentum Onsager matrix in \ttbar-deformed CFTs, which are strongly-interacting field theories, admits such a simple closed form expression. While this result was obtained through the GHD calculation, we confirmed it, up to the second order in $\sigma$, by the conformal perturbative expansion \cite{medenjak2020thermal}.
\vspace{0.2cm}

\noindent {\bf\em $T\bar{T}$-deformed holographic CFTs.}\quad \label{sec:holographic}
The special properties of the $T\bar{T}$-deformation are reflected also in the holographic setup. An irrelevant deformation in general deforms the dual spacetime and destroys the AdS asymptotics. By contrast, the $T\bar{T}$-deformation corresponds to a modification of the boundary conditions at infinity \cite{Guica:2019nzm}. Recall that in holography the leading and subleading terms in the expansion of a bulk field close to the boundary correspond to a source and an expectation value of the operator that is dual to the field. When the field is the graviton, they are the boundary metric and the energy-momentum tensor, respectively. The deformation results in a boundary condition that mixes the two. Remarkably, when the bulk theory is described just by Einstein gravity without extra fields, 
this boundary condition is equivalent to the Dirichlet conditions on a finite-cutoff surface in AdS  (this interpretation, however, makes sense only for one sign of the deformation) \cite{McGough:2016lol}. This is precisely the boundary condition, which we have to consider, since we are only concerned with the dynamics of the energy-momentum tensor. 

Let us discuss the implementation of the partitioning protocol in the scope of the deformed holographic correspondence. 
From the results of the previous part, we expect that we will find a steady state, which is a boosted thermal state. In the CFT case, the dual geometry belongs to the class of Ba\~nados geometries \cite{Banados:1998gg}, with the metric 
\begin{align}\label{Banados}
    ds^2 &= \ell^2 \frac{d\rho^2}{4 \rho^2} + \frac{du dv}{\rho} +{\cal L} (u) \, du^2 +\nonumber\\ &+ \bar {\cal L}(v) \, dv^2 + \rho {\cal L}(u) \bar {\cal L}(v) \, du dv \,.
\end{align}
Here $\rho$ is the radial direction, $u,v$ are the light-cone coordinates at the boundary, $\ell$ is the $AdS$ radius, related to the central charge by $c = \frac{3 \ell}{2 G}$, and the classical gravity regime requires $c \gg 1$. The geometry is parametrized by two arbitrary functions $\cL(u), \bcL(v)$  related to the expectation value of the stress-energy tensor: $\cL = 8 \pi G \ell \langle T_{uu} \rangle, \bcL = 8 \pi G \ell \langle T_{vv}\rangle $. This is the most general solution of the 3d gravity with a flat boundary metric. The thermal state corresponds to the case with constant $\cL, \bcL$ related to the temperature and boost. The corresponding geometry is known as the BTZ black hole.

As shown in \cite{Guica:2019nzm}, the deformed boundary conditions preserve the flatness of the boundary metric, 
implying that the solution can be obtained from \eqref{Banados} with a reparametrization of the boundary coordinates. One can check that, defining $U = u + \mu \int^v \bar{\cal L}(v') dv' \,, \quad V = v + \mu \int^u {\cal L}(u') du' \,, $
the metric \eqref{Banados} in the new coordinates gives an induced metric $dU dV$ at the cutoff surface $\rho=\mu$, so it satisfies the deformed boundary condition. 
This is exactly the same state-dependent change of coordinates that was found in \cite{Dubovsky:2017cnj,Cardy:2019qao}.
The explicit form of the solution can be obtained in the case of our interest, when, as we will see presently,  $\cL,\bcL$ are piece-wise constant and the coordinate change can be inverted, resulting in a deformed metric given by a  non-linear, and non-polynomial expression in $\mu$. 

We can now implement the partitioning protocol, analogously to the case of the undeformed CFTs \cite{Bhaseen:2015aa}. At the initial time the configuration consists of two different boosted thermal states on the left and the right; this state is described by the piece-wise constant functions 
\begin{equation} 
{\cal L}(u) = {\cal L}_L \theta(-u) + {\cal L}_R \theta(u) \,, \ \bar {\cal L}(v) =\bar  {\cal L}_L \theta(-v) + \bar {\cal L}_R \theta(v) \,.
\end{equation}
This solution has a shock-wave singularity at $u=0$ and $v=0$. One can show, inverting the coordinate change, that the trajectory of the shock waves in the new coordinates  $x = (U+V)/2, t=(U-V)/2$ becomes
\begin{equation}
x= - \frac{1+ \mu \bar {\cal L}_L}{1-\mu \bar {\cal L}_L} t \,, \quad x= \frac{1+ \mu {\cal L}_R}{1-\mu {\cal L}_R} t \,.
\end{equation}
The speed of the shock wave agrees with the first-order result of \cite{Bernard_2016} and generalizes it to all orders in $\sigma$. Rewriting the speed as  $v=\sqrt{1+\frac{4 \pi^2 \ell^2 \mu}{\beta^2}}$ we obtain a perfect agreement  with \eqref{effective-v} (we need to relate the cutoff to the deformation parameter: $\mu = - \frac{c}{12 \pi \ell^2} \sigma$ 
which can be derived, {\it e.g.}, by matching the free energy). The case of positive $\mu$, which corresponds to a finite cutoff, leads to superluminal propagation, as observed in \cite{Marolf:2012dr}. 

The solution describing the NESS appearing between the two shock waves is a boosted thermal state with parameters $\cL = \cL_R,\ \bcL = \bcL_L$. Evaluating the holographic energy-momentum tensor we can extract the NESS energy and momentum currents. For the unboosted initial states, with $\cL_L = \bcL_L,\cL_R=\bcL_R$, we find
\begin{align}  \label{JE-hol}
    \langle j_E \rangle_\mathrm{NESS} &=  \frac{1}{8 \pi G \ell} \frac{\cL_L - \cL_R}{1- \mu^2 \cL_L \cL_R} \,, \\
    \langle j_P \rangle_\mathrm{NESS} &= \frac{1}{8 \pi G \ell} \frac{\cL_L + \cL_R + 2 \mu \cL_L \cL_R}{1- \mu^2 \cL_L \cL_R} \,.
\end{align}
These formulae should be compared with \eqref{cicurrent2}; they agree perfectly after converting the parameters in field theory quantities: 
\begin{equation}
    \sqrt{\cL} = \frac{\beta  \left(\sqrt{\frac{4 \pi ^2 l^2 \mu}{\beta ^2}+1}-1\right)}{2 \pi  l \mu } = \frac{\pi \ell}{\tilde{\beta}(\beta,0)} \,.
\end{equation}
Using the linear response theory, the energy and momentum Drude weights can be obtained from the respective NESS currents. 
We find that they are both given by the following expressions involving only thermodynamic quantities: 
\begin{equation}
 D_{EE} =  \frac{\mathtt{e}+\mathtt{p}}{\beta} \,, \quad D_{PP} = \left( \frac{\mathtt{p}}{\mathtt{e}} \right)^2 D_{EE},
\end{equation}
where $\mathtt{e}$ is the energy and $\mathtt{p}$ the momentum density. Note that this expression hold at all orders in the deformation parameter, and can be rewritten as \eqref{Drude-weights}. 

All the holographic predictions so far were in perfect agreement with the GHD results. However, the presence of a non-vanishing diffusion constant for the momentum implies that there should be a broadening of the shock wave into a smooth profile of the densities. This, however, is absent in the gravity solution. 
We can understand the absence of diffusive corrections as a consequence of the classical limit: the result \eqref{LPP} gives an Onsager coefficient at order ${\cal O}(\sigma^2 c^2)$ and correspondingly a diffusion coefficient at ${\cal O}(\sigma^2 c)$. In the classical gravity regime where $c \to \infty$ with $\sigma c$ fixed (in order to keep the cutoff fixed in terms of the AdS radius), the diffusion is suppressed by $1/c$ and therefore the broadening should only be visible from the quantum gravity corrections. Their  computation is in general very hard, but it might be feasible in the 3d gravity. 
 

\vspace{0.2cm}
\noindent {\bf\em Conclusion.}\quad 
In this letter we studied the energy and momentum transport in the $T\bar{T}$-deformed CFTs non-perturbatively, using integrability based generalized hydrodynamics and holography. Despite of limitations of each method, we demonstrated that the exact NESS currents as well as the Drude weights computed by the two approaches match perfectly. This agreement provides a strong verification of the proposed \ttbar-deformed AdS/CFT correspondence from the dynamical point of view. It turns out that the only effect generated by the \ttbar-deformation on the Euler scale, is the deformation of the light-cone velocity, as far as the energy and momentum transport is concerned. Furthermore, we managed to compute the momentum diffusion constant in the \ttbar-deformed CFTs following the recipe of GHD, which is given by a simple universal formula regardless of the underlying CFT. Importantly the energy NESS current we obtained should be in principle measurable experimentally using the noisy thermometry \cite{Jezouin601}. A quantitative analysis of the corrections from \ttbar-deformation in experiment is left for future investigations.

Our findings open a new platform for studying the deformed holographic correspondence on the level of dynamics. In this regard, there are plethora of results that were obtained in the scope of GHD, ranging from entanglement dynamics \cite{Alba7947}, operator spreading \cite{Gopalakrishnan2018}, superdiffusion \cite{Ilievski2018,DeNardis2019Anomalous,PhysRevLett.122.127202,PhysRevLett.125.070601}, integrability breaking \cite{PhysRevB.101.180302,10.21468/SciPostPhys.9.4.044,lopezpiqueres2020hydrodynamics,bastianello2020thermalisation}, which could potentially also be studied by holography. We observed that in \ttbar-deformed CFTs many results can be obtained in a closed form and through independent techniques, such as holography, GHD and perturbed CFT calculations, which makes it a perfect testbed for checking different conjectures obtained in the scope of these approaches.

The findings of the present letter beg many questions also from the physics standpoint. Universality of the momentum diffusion implies that, regardless of the operator content of the underlying CFT, \ttbar-deformed CFTs diffuse in the same way. Such behavior is surprising in view of the recent results on chaos, which does discern between chaotic and non-chaotic CFTs \cite{He2020}. Given the relation between the chaos and diffusion \cite{PhysRevLett.117.091601}, it would be very intersting to understand chaos in \ttbar-deformed CFTs non-perturbatively. 
Here we focused on the energy and momentum transports and the next step would be to study more general setups, pertaining to the other conserved charges, and to the higher-spin analogies of the $T{\bar T}$-deformations.
Finally, we only touched upon a curious connection with the RCA 54, which certainly deserves more attention. In particular, it would be interesting to find RCA, which corresponds to the  \ttbar-deformed AdS exactly.
\vspace{0.2cm}

\noindent {\bf\em Acknowledgements.}\quad
We are indebted to useful comments and discussions with Olalla Castro-Alvaredo, Benjamin Doyon, Monica Guica, Felix Haehl, and Stefano Negro. GP acknowledges the funding from the NYU-PSL research project "Holography and Quantum Gravity" (reference ANR-10-IDEX-0001-02 PSL).  

\vspace{0.2cm}
\bibliographystyle{apsrev4-1}
\bibliography{TC}

\begin{thebibliography}{54}
\bibitem [{com()}]{comment}%
  \BibitemOpen
  \href@noop {} {\ }\bibinfo {note} {Although NLIEs are multi-component in some
  CFTs like minimal models, it does not affect the main idea of the argument
  and yields the same results}\BibitemShut {NoStop}%
\end{thebibliography}
\end{document}